\documentclass[12pt]{article}
\usepackage{amsmath}
\usepackage{amssymb}
\usepackage{verbatim}
\begin{document}

\title{Farrington-Manning in the Extreme Case}
\author{William N. Anderson\footnote{Carpinteria, California, USA. Email: WNilesAnderson@gmail.com. LinkedIn: william-anderson-46384b7}}
\date{May 6, 2022}
\maketitle
\begin{abstract}
The Farrington-Manning method is a common method for evaluating equivalence and non-inferiority of independent proportions.  It is implemented in various software, in particular SAS$^\circledR$ PROC FREQ, and the R$^\circledR$ function farrington.manning(), which is part of the DescrTab2 package. The equations for the estimated proportions can create numerical issues in case both sample proportions equal 1, and neither of these packages will yield an acceptable solution in this special case. In this note we demonstrate a closed form solution for the situation in this and other extreme cases.
\end{abstract}

We assume that the reader has available the original paper of Farrington and Manning \cite{FMformulae}, and we repeat only a minimal amount from the paper. Some text is quoted verbatim from the paper.

\section{The setting}
We consider a comparative binomial trial involving two groups of sizes $N_1, N_2$ in the predetermined ratio $\theta = N_2/N_1$, and independent response variables $r_1 \sim \text{Bi}(N_1, p_1)$ and $r_2 \sim \text{Bi}(N_2, p_2)$. The binomial probabilities $p_1$  and $p_2$ are estimated by $\hat p_1 =r_1/N_1$,
$\hat p_2 =r_2/N_2$. The true difference is $s_T = p_1 - p_2$, estimated by $s = \hat p_1 - \hat p_2$. In the equivalence or non-inferiority setting the null hypothesis is $p_1 - p_2 = s_0$, where $s_0$ is a predetermined value, often called the {\em non-inferiority margin}. 

The Farrington-Manning method is to compute maximum likelihood estimates  $\tilde p_{1D}$ and  $\tilde p_{2D}$ for  $p_1$  and $p_2$, under the constraint that $\tilde p_{1D} - \tilde p_{2D} = s_0$. These estimates will typically be different from $\hat p_1$  and $\hat p_2$. The method is described in the original paper~\cite{FMformulae}, and in many other places, including section 11.2.3 of Rothmann et al.~\cite{RothmannDesign}. This method is implemented in many statistical software packages, including SAS$^\circledR$ PROC FREQ,  and the R$^\circledR$ function farrington.manning(), which is part of the DescrTab2 package. \footnote{The farrington.manning() function was written by Kevin Kunzmann, and the DescrTab2 package is maintained by Jan Meis.  As of the date of this note, Version 2.1.9 of the package, dated January 20, 2022,  is the most current. } \footnote{Thanks to Jan Meis for many helpful comments relating to the package, and to the issues described here.}

There is no restriction on the sign of $s_0$, but, of course, $s_0$ must have a value suitable for the difference of proportions. Reading between the lines, it appears that Farrington and Manning had in mind positive $s_0$. In practice, one can arrange this choice by choosing the groups appropriately. The farrington.manning() function allows for negative $s_0$; SAS PROC FREQ does not. 

Then the test statistic is 
\begin{equation}
z_D = \hat p_1 - \hat p_2 - s_0.
\end{equation}

For large $N$,  $z_D$ is approximately Normally distributed. Under the null hypothesis, the variance
of $z_D$ is estimated by:
\begin{align}\label{nullvarianceequation}
\hat v_0 &= \nonumber \tilde p_{1D} \tilde q_{1D}/N_1 + \tilde p_{2D} \tilde q_{2D}/N_2\\
             &= \left[\tilde p_{1D} \tilde q_{1D} + \tilde p_{2D} \tilde q_{2D} \theta \right]/N_1
\end{align}
where $\tilde p_{1D}$ and $\tilde p_{2D}$ are maximum likelihood estimators for $p_1$ and $p_2$ under the null hypothesis and $\tilde q_{1D} = 1 - \tilde p_{1D}$, $\tilde q_{2D} = 1 - \tilde p_{2D}$. The null hypothesis may then be tested by referring the statistic:
\begin{equation}\label{zstatistic}
z = (\hat p_1 - \hat p_2 - s_0) / \sqrt{\hat v_0}
\end{equation}
to the standard normal distribution.

In the general case, the estimates $\tilde p_{1D}$ and $\tilde p_{2D}$ are obtained by solving a cubic equation, as will be seen below. 

However, this procedure has potential numerical issues when both $\hat p_1 = 1$ and  $\hat p_2 = 1$.  This situation may seem unlikely in an actual trial, but it can occur when one is doing a simulation with assumed probabilities $p_1$  and $p_2$ both close to 1.  In this situation the SAS PROC FREQ simply refuses to compute the risk difference, and the R function farrington.manning() may cause an error condition.  

\section{Computing the maximum likelihood estimators}\label{maximumlikelihoodsection}

The estimate $\tilde{p}_{1D}$ is obtained by solving the maximum likelihood equation

\begin{equation}\label{maximumlikelihoodequation} 
 ax^3 + bx^2 + cx + d = 0
  \end{equation}

with 

\begin{align*}
a &= 1 + \theta \\
b &= - [1 + \theta + \hat p_1 + \theta \hat p_2 + s_0(\theta + 2)] \\
c &= s_0^2 + s_0(2 \hat p_1 + \theta + 1) + \hat p_1 + \theta \hat p_2 \\
d &= -\hat p_1 s_0 (1 + s_0)
\end{align*}

Any method for solving a cubic equation could in principle be used to compute  $\tilde{p}_{1D}$, but one must be careful to choose the root that maximizes the underlying likelihood function. Some considerations on this point are given in section \ref{SolutionConsiderations} below.

\bigskip

In the special case of interest here, both $\hat p_1 = 1$ and $\hat p_2 = 1$. The coefficients of the cubic equation \eqref{maximumlikelihoodequation} take a simpler form, and the equation can be factored.  
\begin{align}\label{case0equation}
\nonumber ax^3 + bx^2 + cx + d &= (1 + \theta)x^3 \\
\nonumber &\quad-  [2 + 2\theta  + s_0(\theta + 2)]x^2 \\
\nonumber &\quad+ [s_0^2 + s_0(3 + \theta) + 1 + \theta]x \\
\nonumber &\quad- s_0(1 + s_0) \\
&= (x - 1)(x - (1 + s_0))((1 + \theta)x - s_0).
\end{align}

The cubic has three real roots, and the choices for $\tilde p_{1D}$  and  $\tilde p_{2D}$ are given in table \ref{case1roots} below.

\begin{equation} \label{case0roots}
\begin{bmatrix} \tilde p_{1D} \\ \tilde p_{2D} \end{bmatrix} =
\begin{bmatrix} 1 & 1 + s_0 & s_0/(1 + \theta) \\ 1 - s_0 & 1 & -s_0\theta/( 1 + \theta) \end{bmatrix}
\end{equation}

 \begin{itemize}
 \item If $s_0 > 0$, only $\tilde p_{1D} = 1$ and $\tilde p_{2D} = 1 - s_0$ give valid proportions for $\tilde p_{1D}$ and $\tilde p_{2D}$.
  \item If $s_0 <0$, only $\tilde p_{1D} = 1 + s_0$ and $\tilde p_{2D} = 1$ give valid proportions for $\tilde p_{1D}$ and $\tilde p_{2D}$. 
 \end{itemize}
 
 An alternative method for finding the estimates is to note that the likelihood function is monotonic increasing in  $\tilde p_{1D}$ and $\tilde p_{2D}$. Accordingly the maximum likelihood will be obtained when these two are as large as possible. That is when one of the $\tilde p_{iD}$ is 1, and the other is $1 - |s_0|$.

\section{Hypothesis tests and confidence intervals}\label{hypothesissection}

Using the general variance equation~\eqref{nullvarianceequation}, the variance then becomes 
\begin{equation}\label{nullvariancespecial}
\hat v_0 = \begin{cases} s_0(1 - s_0)/N_2 & s_0 > 0 \\
-s_0(1 + s_0)/N_1 & s_0 < 0
\end{cases}
\end{equation}

Using this variance and the test statistic of \eqref{zstatistic} we have for $s_0 > 0$

\begin{align*}
z &= \nonumber (\hat p_1 - \hat p_2 - s_0) /\sqrt{\hat v_0} \\
&= -s_0/\sqrt{ s_0(1 - s_0)/N_2} \\
&= -\sqrt{N_2 s_0/(1 - s_0)}.
\end{align*}

Similarly, for $s_0 < 0$
\begin{align*}
z &= \nonumber (\hat p_1 - \hat p_2 - s_0) /\sqrt{\hat v_0} \\
&= -s_0/\sqrt{ -s_0(1 + s_0)/N_1} \\
&= \sqrt{-N_1 s_0/(1 + s_0)}.
\end{align*}

The two-sided confidence limits for the difference are obtained by inversion of the two-sided hypothesis test  For the general Farrington-Manning case the computations require an iterative solution, because there is no simple formula relating the $z$-statistic to $s_0$. For the special case here limits are simply obtained by setting $z$ to the critical values, and solving for $s_0$. 

The confidence interval is 
 
\begin{equation*}
\left(\dfrac{-z_\alpha^2}{N_1 + z_\alpha^2},  \dfrac{z_\alpha^2}{N_2 + z_\alpha^2}\right)
\end{equation*}

As a practical matter, if one really had data this extreme, exact methods would likely be used for evaluating hypothesis tests and computing confidence limits. The intention here is simply to give reasonable limits that can be used to let a simulation proceed.  

\section{Other extreme cases}

A number of other extreme cases can be considered in the same manner. Here we give only minimal details, leaving the rest to the reader.
 
\bigskip\par\noindent{\em Suppose  $\hat p_1 = 0$ and $\hat p_2 = 0$.} 

The maximum likelihood equation~\eqref{maximumlikelihoodequation} factors
\begin{align}\label{case1equation}
\nonumber ax^3 + bx^2 + cx + d &= (1 + \theta)x^3 \\
\nonumber &\quad-  [1 + \theta + s_0(\theta + 2)]x^2 \\
\nonumber &\quad+ [s_0^2 + s_0( \theta + 1)  ]x \\
\nonumber &\quad- 0 \\
&= x(x - s_0)((1 + \theta)x - (1+ \theta + s_0)).
\end{align}

The possibilities for $\tilde p_{1D}$  and  $\tilde p_{2D}$ are given in table \eqref{case1roots} below. The appropriate choice depends on the sign of $s_0$. 

\begin{equation} \label{case1roots}
\begin{bmatrix} \tilde p_{1D} \\ \tilde p_{2D} \end{bmatrix} =
\begin{bmatrix} 0 & (1 + \theta + s_0)/(1 + \theta) & s_0 \\ -s_0 & (1 + \theta - s_0\theta)/(1 + \theta) & 0\end{bmatrix}
\end{equation}

\begin{itemize}
 \item If $s_0 > 0$, only $\tilde p_{1D} = s_0$ and $\tilde p_{2D} = 0$ give valid proportions for $\tilde p_{1D}$ and $\tilde p_{2D}$.
 \item If $s_0 <0$, only $\tilde p_{1D} = 0$ and $\tilde p_{2D} = -s_0$ give valid proportions for $\tilde p_{1D}$ and $\tilde p_{2D}$. 
 \end{itemize}
 
The remaining computations are highly similar to those above. 
 
\bigskip\par\noindent{\em Suppose  $\hat p_1 = 0$ and $\hat p_2 = 1$.} The maximum likelihood equation~\eqref{maximumlikelihoodequation} factors

\begin{align}\label{case2equation}
\nonumber ax^3 + bx^2 + cx + d &= (1 + \theta)x^3 \\
\nonumber &\quad-  [1 + 2\theta + s_0(\theta + 2)]x^2 \\
\nonumber &\quad+ [s_0^2 + s_0(\theta + 1) + \theta]x \\
\nonumber &\quad- 0 \\
&= x(x - (1 + s_0))((1 + \theta)x - (\theta + s_0)).
\end{align}

The possibilities for $\tilde p_{1D}$  and  $\tilde p_{2D}$ are given in table \eqref{case2roots} below. The appropriate choice depends on the sign of $\theta + s_0$.

\begin{equation} \label{case2roots}
\begin{bmatrix} \tilde p_{1D} \\ \tilde p_{2D} \end{bmatrix} =
\begin{bmatrix} 0 & (\theta + s_0)/(1 + \theta) & 1 + s_0 \\ -s_0 & (\theta - s_0\theta)/(1 + \theta) & 1\end{bmatrix}
\end{equation}

\begin{itemize}
 \item If $\theta + s_0 > 0$, the choice $\tilde p_{1D} =  (\theta + s_0)/(1 + \theta)$ and $\tilde p_{2D} = (\theta - s_0\theta)/(1 + \theta)$ gives the maximum likelihood estimator. (If additionally $s_0 < 0$, all three choices give valid proportions, but the other two choices give minima).
 \item If $\theta + s_0 = 0$, the choice  $\tilde p_{1D} =  0$ and $\tilde p_{2D} =-s_0$ is a double root, and gives the maximum. 
   \item If $\theta + s_0 <0$, the choice $\tilde p_{1D} =  0$ and $\tilde p_{2D} =-s_0$ gives the maximum. (The choice  $\tilde p_{1D} =  1 + s_0$ and $\tilde p_{2D} = 1$ also gives valid proportions, but this is a minimum.) 
 \end{itemize}
 
 The remaining computations proceed as above. The upper confidence limit for the difference will be 1; we have not found a simple form for the lower confidence limit. 
 
\bigskip\par\noindent{\em Suppose  $\hat p_1 = 1$ and $\hat p_2 = 0$.}  The maximum likelihood equation~\eqref{maximumlikelihoodequation} factors

\begin{align*}
ax^3 + bx^2 + cx + d &= (1 + \theta)x^3 \\
&\quad-  [2 + \theta + s_0(\theta + 2)]x^2 \\
&\quad+ [s_0^2 + s_0(\theta + 3) + 1]x \\
&\quad- s_0(1 + s_0)\\
&= (x - 1)(x - s_0)((1 + \theta)x - (1 + s_0)).
\end{align*}

The possibilities for $\tilde p_{1D}$  and  $\tilde p_{2D}$ are given in table \eqref{case3roots} below. The appropriate choice depends on the sign of $1 - s_0\theta$.

\begin{equation} \label{case3roots}
\begin{bmatrix} \tilde p_{1D} \\ \tilde p_{2D} \end{bmatrix} =
\begin{bmatrix} s_0 & (1 + s_0)/(1 + \theta) & 1 \\ 0 & (1 - s_0\theta)/(1 + \theta) & 1 - s_0\end{bmatrix}
\end{equation}

\begin{itemize}
 \item If $1 - s_0\theta > 0$, the choice $\tilde p_{1D} =  (1 + s_0)/(1 + \theta)$ and $\tilde p_{2D} = (1 - s_0\theta)/(1 + \theta)$ gives the maximum likelihood estimator. (If additionally $s_0 > 0$, all three choices give valid proportions, but the other two choices give minima).
 \item If $1 - s_0\theta = 0$, the choice  $\tilde p_{1D} =  s_0$ and $\tilde p_{2D} =0$ is a double root, and gives the maximum. 
  \item If $1 - s_0\theta <0$, the choice $\tilde p_{1D} =  0$ and $\tilde p_{2D} =-s_0$ gives the maximum. (The choice  $\tilde p_{1D} =  1$ and $\tilde p_{2D} = 1 - s_0$ also gives valid proportions, but this is a minimum.) 
 \end{itemize}
 
 The remaining computations proceed as above. The lower confidence limit for the difference will be -1; we have not found a simple form for the upper confidence limit. 

 \bigskip\par\noindent{\em Suppose $s_0 = 0$.} This is in effect a standard superiority test, but there seems to be no mathematical reason not to compute using the methods above. In this case the special form of the equation becomes.
 
 \begin{align*}
ax^3 + bx^2 + cx + d &= (1 + \theta)x^3 \\
&\quad-  [1 + \theta + \hat p_1 + \theta \hat p_2]x^2 \\
&\quad+ [\hat p_1 + \theta \hat p_2]x \\
&\quad- 0\\
&= x(x - 1)((1 + \theta)x - (\hat p_1 + \theta\hat p_2)).
\end{align*}

The desired solution is $x = (\hat p_1 + \theta\hat p_2)/(1 + \theta) = (r_1 + r_2)/(N_1 + N_2)$.  The resulting statistical test is algebraically identical to the chi-squared test without the continuity correction.  
 
\section{Solution Considerations} \label{SolutionConsiderations}

The confidence intervals above are computed by inverting the two-sided hypothesis test.  We note that the intervals using the farrington.manning() function are obtained in this manner, while those in SAS PROC FREQ are not. Various other methods of computing confidence intervals are discussed in the textbooks of Newcombe~\cite{NewcombeConfidence} and Rothmann et al~\cite{RothmannDesign}.  In reading any of the references it is vital to check on the definition of $s_0$; in some cases the sign is reversed from the original paper. 

The Farrington-Manning paper \cite{FMformulae} states without proof that the cubic equation \eqref{maximumlikelihoodequation} has a unique solution in $(s_0, 1)$. 
\begin{itemize} 
\item For $s_0 > 0$, the interval $(s_0, 1)$ makes sense, since $\tilde{p}_{2D} = \tilde{p}_{1D} - s_0$ must be a proportion.  
\item But if $s_0 < 0$, the appropriate interval is $(0, 1 + s_0)$, because $\tilde{p}_{2D} > \tilde{p}_{1D}$. This situation is why it seems that the authors had in mind $s_0 > 0$.  
\end{itemize}

The paper \cite{FMformulae} uses the cubic formula to solve~\eqref{maximumlikelihoodequation}. The particular form of the cubic formula is the trigonometric form, which is appropriate for the case when the cubic has three real roots. Only one of the three roots is given by the formula used in the paper. It appears that the this form of the cubic formula is appropriate, and that the chosen root is the correct one, even in the extreme cases.  However, the paper  \cite{FMformulae}  does not give any proofs in this regard.  In this note we simply accept the situation, and have made no attempt to present proofs. This use of the cubic formula is implemented in both SAS PROC FREQ, and the R function farrington.manning(). It is straightforward to program the formula in Excel.  

Numerical problems with the formulas occur in two different situations:
\begin{itemize}
\item The cubic formula involves computing an arc cosine. If there is a double root, the numeric argument for the arc cosine function may be exactly 1. But roundoff can cause the computed argument to be slightly greater than 1. One can fix the problem using the formulas for $\tilde{p}_{1D}$ and $\tilde{p}_{1D}$ above. Alternatively one could check the argument of the arc cosine function, and set arguments microscopically greater than 1 to exactly 1.

An example would be  when $\hat p_1 = 0$, $\hat p_2 = 1$, $\theta = 2$, and $s_0 =  0.5$. In computation for the farrington.manning() function roundoff causes the argument of the acos() function to be microscopically greater than 1, which produces an error condition.  One could fix the situation by checking the argument before the acos() function is used. Excel does not generate an error for this example, nor does SAS. 

\item  Inverting the computation to find the confidence limits involves solving a non-linear equation over a defined range.  Too small a range can result in missing the root. This situation can arise in the case $\hat p_1 = 0$ and $\hat p_2 = 1$, or vice versa. In these cases one confidence limit is obvious, and it may be reasonable to write special code to compute the other limit. As mentioned above, if real data produced this situation one would likely use exact methods, and the formulas here would be irrelevant.

\item Jan Meis~\cite{Miesprogram} has modified the farrington.manning() function to incorporate both suggestions in this paragraph. Those suggestions might appear in a later version of the DescrTab2 package.
\end{itemize}
\bigskip
In the examples above the closed form values for $\tilde p_{1D}$ and $\tilde p_{2D}$ were obtained by factoring the cubic equation~\eqref{maximumlikelihoodequation}. One could speculate as to whether such factoring can always be done. For an example to the contrary, suppose that $\hat p_1 = 1/3$, $\hat p_2 = 1/2$, $s_0 = 1/2$, and $\theta = 2/3$.  Then the equation becomes 
\begin{align}\label{norationalroot}
\nonumber (5/3)x^3 - (11/3)x^2 + (25/12) x - (3/12) &= 0\\
20 x^3 - 44x^2 + 25x - 3 &= 0
\end{align}
Any rational root of the form $m/n$ must have $m \mid 3$ and $n \mid 20$. There are only a finite number of cases to check, and none yield a rational root.  Since~\eqref{norationalroot} is a cubic, it must therefore be irreducible over the rationals, and the full complications of cubic fields enter into the solution.

\nocite{FMformulae}
\nocite{Rcitation}
\nocite{DescrTab2citation}
\nocite{SASSTAT}
\nocite{Miesprogram}
\nocite{NewcombeConfidence}
\nocite{RothmannDesign}
\bibliographystyle{plain}
\bibliography{FM.bib}

\end{document}